\documentclass{article}
\PassOptionsToPackage{table, dvipsnames}{xcolor}
\usepackage{arxiv}
\usepackage[utf8]{inputenc} % allow utf-8 input
\usepackage[T1]{fontenc}    % use 8-bit T1 fonts
\usepackage{url}            % simple URL typesetting
\usepackage{nicefrac}       % compact symbols for 1/2, etc.
\usepackage{soul}
\usepackage{fontawesome}
\usepackage[many]{tcolorbox}
\usepackage[table]{xcolor}
\usepackage{booktabs} % for professional tables
\usepackage{lipsum}
\usepackage{xspace}
\usepackage{enumitem}
\usepackage{natbib}
\usepackage{xcolor}
\definecolor{darkblue}{RGB}{44,62,80}
\definecolor{CalGoldHex}{RGB}{253, 181, 21} % FDB515
\definecolor{gold}{RGB}{149, 113, 30} % 95711E
\usepackage{tikz}
\usetikzlibrary{arrows.meta,positioning,fit,calc,shapes.multipart}

\usepackage[breaklinks=true,colorlinks,bookmarks=false,citecolor=blue,linkcolor=darkblue]{hyperref}
\usepackage{tcolorbox}
\tcbuselibrary{skins,breakable}
\usepackage{lmodern}

\newtcolorbox{promptbox}[1]{
  breakable,
  enhanced,
  sharp corners,
  colback=gray!3,
  colframe=black!12,
  coltitle=black,
  title=\textbf{#1},
  fonttitle=\bfseries,
  boxrule=0.6pt,
  left=10pt,right=10pt,top=10pt,bottom=10pt,
  attach boxed title to top left={yshift*=-2mm, xshift=2mm},
  boxed title style={
    colback=black!5,
    colframe=black!12,
    sharp corners,
    boxrule=0.6pt,
    top=4pt,bottom=4pt,left=6pt,right=6pt
  },
  fontupper=\footnotesize\ttfamily
}

\usepackage{enumitem}
\usepackage{doi}

\usepackage{amsmath}
\usepackage{amssymb}
\usepackage{mathtools}
\usepackage{amsthm}

\usepackage[capitalize,noabbrev]{cleveref}

\usepackage{fourier}

\usepackage[textsize=tiny]{todonotes}

\usepackage{tabularx}

\title{The Role of Computing Resources in Publishing Foundation Model Research}

\newcommand*{\affmark}[1][*]{\textsuperscript{\textnormal{#1}}}

\author{
\textbf{Yuexing Hao} \affmark[1, 2]$^{\dagger}$,
\textbf{Yue Huang} \affmark[3]$^{\dagger}$, 
\textbf{Haoran Zhang} \affmark[1], 
\textbf{Chenyang Zhao} \affmark[4], 
\textbf{Zhenwen Liang} \affmark[3],\\ 
\textbf{Paul Pu Liang} \affmark[1], 
\textbf{Yue Zhao} \affmark[6], 
\textbf{Lichao Sun} \affmark[5], 
\textbf{Saleh Kalantari} \affmark[2],
\textbf{Xiangliang Zhang} \affmark[3],
\textbf{Marzyeh Ghassemi} \affmark[1] \\ 
\affmark[1]EECS, MIT, Cambridge, 02135, USA.\\
\affmark[2]Cornell University, Ithaca, 14850, USA.\\
\affmark[3]CSE, University of Notre Dame, South Bend, 46556, USA.\\
\affmark[4]Computer Science Department, University of California, Los Angeles, 90095, USA.\\
\affmark[5]Computer Science Department, Lehigh University, Bethlehem, 18015, USA.\\
\affmark[6]School of Advanced Computing, University of Southern California, Los Angeles, 90007, USA.\\
\small$^\ast$Corresponding author. Email: yuexing@mit.edu \\
\small$^\dagger$These authors contributed equally to this work.\\
\\
}

\renewcommand{\shorttitle}{The Role of Computing Resources in Publishing Foundation Model Research}
\usepackage{threeparttable}
\usepackage{pythonhighlight}
\usepackage{booktabs}
\usepackage{graphicx}
\usepackage{booktabs}
\usepackage{caption}
\usepackage{subcaption}
\usepackage{fourier}
\usepackage{multirow}
\usepackage{comment}
\usepackage{wrapfig}
\usepackage{array}
\newcolumntype{C}[1]{>{\centering\arraybackslash}m{#1}}
\newcolumntype{L}[1]{>{\raggedright\arraybackslash}m{#1}}

\usepackage{color}
\definecolor{darkblue}{RGB}{44,62,80}
\definecolor{CalGoldHex}{RGB}{253, 181, 21} % FDB515
\definecolor{gold}{RGB}{149, 113, 30} % 95711E
\usepackage{tcolorbox} 

\tcbset{
  definitionbox/.style={
    colback=green!5!white,
    colframe=black!50,
    boxrule=0.5pt,
    sharp corners,
    left=1.5mm,
    right=1.5mm,
    top=1mm,
    bottom=1mm,
  }
}

\definecolor{deepred}{rgb}{0.631,0.102,0.102}
\definecolor{amethyst}{rgb}{0.6, 0.4, 0.8}
\definecolor{darkgreen}{rgb}{0.3,0.7,0.3}
\definecolor{salmon}{RGB}{241, 150, 141}

\begin{document}
\maketitle
\begin{abstract}
Cutting-edge research in Artificial Intelligence (AI) requires considerable resources, including Graphics Processing Units (GPUs), data, and human resources. In this paper, we evaluate of the relationship between these resources and the scientific advancement of foundation models (FM). We reviewed 6517 FM papers published between 2022 to 2024, and surveyed 229 first-authors to the impact of computing resources on scientific output. We find that increased computing is correlated with national funding allocations and citations, but our findings don't observe the strong correlations with research environment (academic or industrial), domain, or study methodology. We advise that individuals and institutions focus on creating shared and affordable computing opportunities to lower the entry barrier for under-resourced researchers. These steps can help expand participation in FM research, foster diversity of ideas and contributors, and sustain innovation and progress in AI. The data will be available at: \url{https://mit-calc.csail.mit.edu/}.
\end{abstract}

\begin{center}
\end{center}

\section{Introduction}

Artificial Intelligence (AI) and machine learning (ML) models have made stark advances in the past three years, fueled by the development of foundation models (FM) trained on large-scale multimodal data. Following the public release of several successful FMs (\cite{open_ai_team_introducing_2022, brown_language_2020, bommasani_opportunities_2022-1}), FMs such as large language models (LLMs) and vision language models (VLMs) have bridged vision, language, and other modalities. 
In many Computer Science subfields such as Natural Language Processing (NLP) and Computer Vision (CV), FMs have demonstrated strong compositional performance and generalization capabilities (\cite{awais_foundation_2025, gunter_apple_2024}), emerging as widely-used tools (\cite{bommasani_opportunities_2022-1}) that provide a flexible backbones for innovation in other fields (\cite{moor_foundation_2023, sartor_neural_2025, firoozi_foundation_2024}).

Conducting FM research requires significant data, computing, and human resources (\cite{cottier_rising_2024, maslej_ai_2024, crawford_generative_2024}). A central concern in the field is whether greater access to such resources directly translates into more impactful research outcomes (\cite{acemoglu_simple_2024, dodge_show_2019, open_ai_team_ai_2018}), such as more research publications, or higher citation counts (\cite{sinclair_fifty_2023, anjum_retrospective_2019}). The answer to this question has important implications for how resources are allocated, which research directions are prioritized, and how equitable participation in FM research can be ensured.
However, the cost of research is often difficult to quantify due to lack of uniform disclosure on resource distribution (\cite{bommasani_considerations_2024}). 
Absent widespread disclosure, funding is perhaps most easily characterized in the concrete cost of purchasing or renting hardware (e.g., computing clusters, or chips), through there are also software, cloud storage services, and specialized software platform costs. %the interplay between academic and industry stakeholders, and the distinction between open-source and closed-source models 
GPUs are a particularly salient metric for examination as they are a tightly controlled resource available in fixed quantities (\cite{owens_gpu_2008, nickolls_gpu_2010}), and there is often intense competition for their purchase (\cite{khandelwal_100k_2024, kudiabor_ais_2024}). 

In this paper, we examine the relationship between hardware  resources and publications in selective AI/ML computer science conferences. We focus on benchmarking two measurements of computing power - the number of graphical processing units (GPUs) and teraflops (TFLOP) - and relate those to data from over 34,828 accepted papers between 2022 and 2024. 

We identify 5,889 FM papers published and find that greater GPU access is associated with higher acceptance rates in eight conference venues, as well as with higher citation counts. While this relationship is likely influenced by other factors (i.e. institutional resources, reputation, and researcher networks), the trend demonstrates the potential role of computing resources in shaping FM research visibility and impact. We also survey 229 authors of 312 papers, to identify any discrepancies between reported usage and true usage of resources. We find that: (1) the majority of foundation model papers are authored by researchers in academia (4,851 papers) compared to those from industry (1,425 papers); (2) most papers used open-source models (i.e. Llama), followed by closed-source models (i.e. GPT); and (3) GPU information is rarely disclosed in the manuscript, demonstrating the need for standardized reporting of computing resources to improve transparency and reproducibility.

\vspace{-7pt}
\section{Identifying Computing Resources}
\subsection{LLM-extracted Publication Data}

\begin{figure}[t]
	\centering
\includegraphics[width=0.75\linewidth]{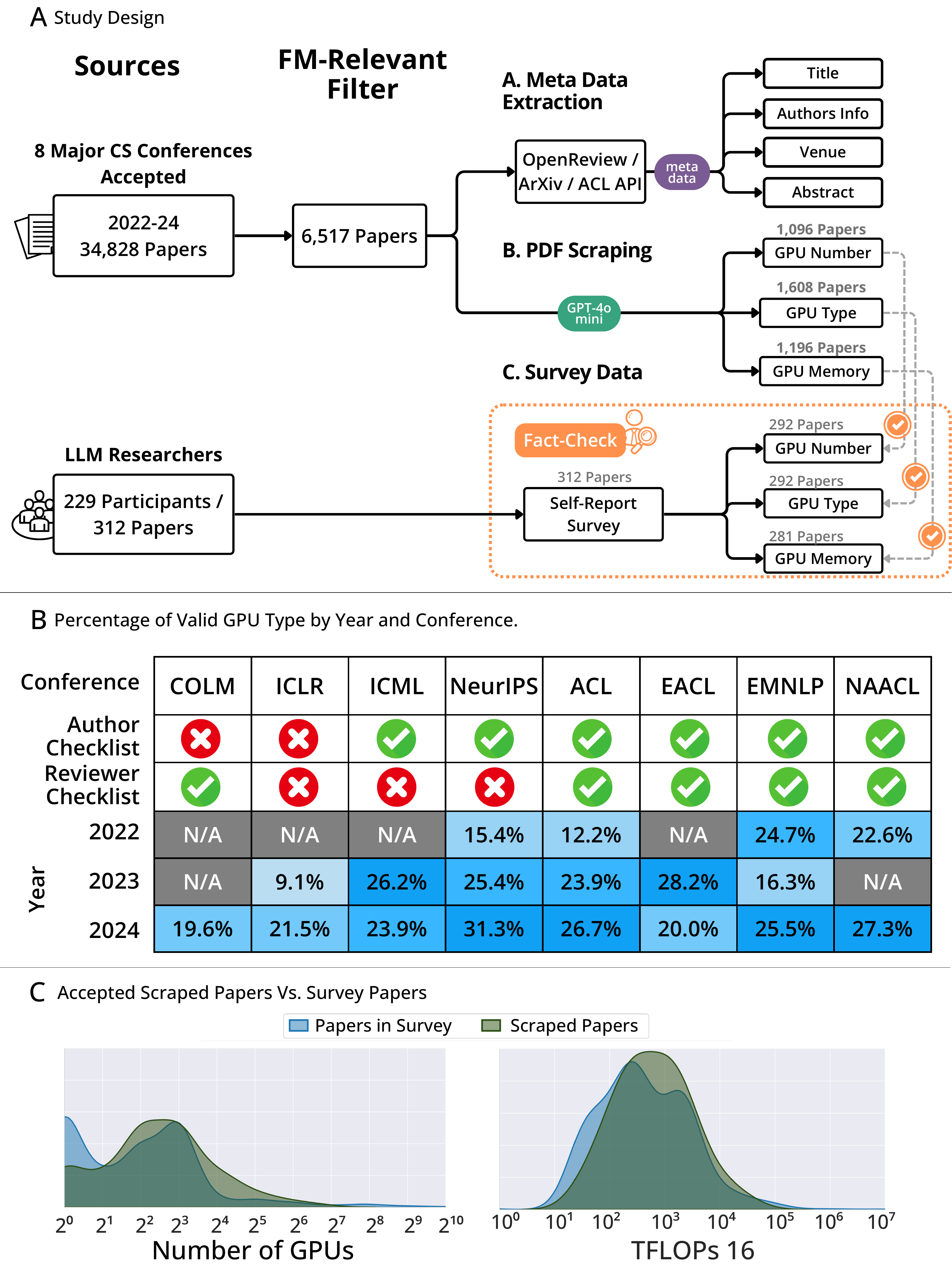} 
	\caption{\textbf{Study Design and Data Collection.}}
	% 	(A) We initially collected 34,828 papers from eight major computer science conferences, from which 6,517 FM-related papers were retrieved using the OpenReview API and the ACL ARR platform. Then we extracted their pdfs with GPU related information. In the survey study, we recruited 229 first-author FM researchers, representing 312 papers in total, to participate in our survey. Participants provided self-reported responses regarding computing resources when such information was not documented in their publications. Dotted boxes indicate potentially unavailable information. (B) Percentage of valid GPU type by year and conference, and whether the conference author and reviewer checklists contain related guidelines to report these computing resource usage. Noted that ARR conferences (including ACL, EACL, EMNLP, and NAACL) used the same author and reviewer checklists with slightly modifications based on each conference's requirements. (C) GPU Usage and TFLOPS 16 Between GPT-4o Scraped and Self-Reported Survey Data.}
	\label{fig:study_design} 
    \vspace{-20pt}
\end{figure}

We collect accepted papers from eight top machine learning conferences between 2022-2024 that were available as of March 2025 (Figure \ref{fig:study_design} panel (A)). We access papers via the OpenReview API (NeurIPS, ICLR, ICML, COLM, and EMNLP 2023) and the Association for Computational Linguistics (ACL) Rolling Review (ARR) platform (ACL, NAACL, EACL, EMNLP 2022 and 2024). %All eight conferences follow a double-blind review process, in which both authors and reviewers remain anonymous throughout the review period. 
We identify FM-related papers using keywords in the titles or abstracts. A total of 5,889 accepted papers were identified as being FM-relevant from 34,828 total papers. We also collect rejected and withdrawn FM-related ICLR papers - a total of 613 rejected or withdrawn ICLR papers in the same period for comparison. A full description of this process can be found in the appendices. 

We collect structured information using system APIs from all 5,889 accepted papers, including IDs of the articles, titles, author details (name(s), number of authors and affiliated institutions), publication details (such as year, venue and accept / reject status, paper link, reviews and abstracts (Figure \ref{fig:study_design}).
We use the GPT-4o mini (\cite{openai_gpt-4_2024}) LLM to collect information missing from the system API by processing article PDFs to extract the senior author's affiliation, GPU usage, data set descriptions, and funding information. In our analyses, we attribute the paper's affiliation to the senior author listed. For papers involving multiple institutions or countries, we use only the first institution and country listed for the senior author. If a listed institution has multiple locations, e.g., Google Research, we map this to the publicly listed country headquarters. After compiling the titles and abstracts of the papers, we used GPT-4o Mini to classify each paper into three categories: Domain, Phase, and Method. Definitions for these categories are provided in Table \ref{definition_table}. 

We initially collected 34,828 papers from eight major computer science conferences, from which 6,517 FM-related papers were retrieved using the OpenReview API and the ACL ARR platform. Then we extracted their pdfs with GPU related information. In the survey study, we recruited 229 first-author FM researchers, representing 312 papers in total, to participate in our survey. Participants provided self-reported responses regarding computing resources when such information was not documented in their publications. Dotted boxes indicate potentially unavailable information (Figure \ref{fig:study_design} panel (A)). Percentage of valid GPU type by year and conference, and whether the conference author and reviewer checklists contain related guidelines to report these computing resource usage is presented in Figure \ref{fig:study_design} panel (B). Noted that ARR conferences (including ACL, EACL, EMNLP, and NAACL) used the same author and reviewer checklists with slightly modifications based on each conference's requirements. Figure \ref{fig:study_design} panel panel (C) shows the GPU Usage and TFLOPS 16 between GPT-4o scraped and self-reported survey data.

To ensure the accuracy of the extracted GPU information, two FM researchers independently checked 312 papers and compared to GPT-4o mini in a blinded setting. We cross-referenced the information extracted by GPT-4o mini, the researchers' annotations, and the GPU counts self-reported by first-author participants.
When authors of the 312 papers were surveyed, 288 papers self-reported GPU numbers, 292 papers self-reported GPU types, and 281 papers self-reported GPU hours. 24 papers used non-GPU resources (e.g. TPU, NPU, CPU). However, FM researchers found that only 172 papers' PDFs contained GPU numbers, 141 papers' PDFs contains GPU types, and 249 papers' PDFs contains GPU hours. GPT-4o mini was able to scrape GPU numbers from just 116 papers, showing a 59.7\% gap compared to author reports. The missing rates were also high for GPU type (48.3\%) and GPU hours (88.6\%).

\begin{table*}[h!]
\centering
\renewcommand{\arraystretch}{1.45}
\rowcolors{2}{white}{gray!10}
\setlength{\tabcolsep}{6pt}

\begin{tabularx}{\textwidth}{>{\raggedright\arraybackslash}p{0.1\textwidth}X}
\toprule[1pt]
\textbf{Category} & \textbf{Justification and Distinction} \\
\midrule

\textbf{Domain} &
\textbf{Natural Language Processing (NLP)}: Involves text and language data, addressing tasks such as translation, summarization, question answering, and dialogue generation. NLP remains one of the most established FM domains (e.g., GPT, BERT). \\

& \textbf{Computer Vision (CV)}: Deals with visual data, including image classification, object detection, segmentation, and video understanding. Vision FMs handle diverse modalities and architectures (e.g., ViT, SAM). \\

& \textbf{Data Mining}: Applies FMs to structured or semi-structured data for pattern discovery, anomaly detection, and recommendation systems. \\

& \textbf{Security}: Focuses on applications in threat detection, adversarial robustness, privacy protection, and secure model deployment, often intersecting with adversarial ML and interpretability. \\

& \textbf{Explainable AI (XAI)}: Aims to improve model transparency by probing FM reasoning, generating explanations, and designing interpretable architectures. \\

& \textbf{Others}: Includes interdisciplinary or emerging areas such as systems optimization (e.g., FM efficiency, scaling laws) and human-computer interaction. \\

\midrule
\textbf{Phase} &
\textbf{Pre-trained}: Refers to large-scale training on general-purpose datasets (e.g., web-scale or multi-modal corpora), establishing foundational model capabilities. \\

& \textbf{Post-trained}: Involves fine-tuning, domain adaptation, or task-specific refinement using smaller or labeled datasets to narrow the model's focus. \\

& \textbf{Inference}: Represents the deployment stage, emphasizing efficiency, robustness, safety, and real-world alignment. \\

\midrule
\textbf{Method} &
\textbf{Algorithm}: Introduces new architectures, optimization strategies, or training paradigms (e.g., sparse attention, contrastive learning) that enhance FM performance or capabilities. \\

& \textbf{Dataset/Benchmark}: Develops datasets or evaluation suites (e.g., BIG-Bench, HELM) to measure FM performance, standardize comparisons, and expose limitations. \\

& \textbf{Empirical Study}: Conducts systematic analyses of existing models to examine scaling laws, emergent behaviors, or social biases, often revealing critical limitations. \\

& \textbf{Toolkit}: Builds open-source frameworks or libraries (e.g., Hugging Face Transformers, LangChain) that promote usability, reproducibility, and extensibility in FM research. \\

\bottomrule[1pt]
\end{tabularx}
\caption{\textbf{Justification of the three classification dimensions used to categorize foundation model (FM) research papers.}}
\label{definition_table}
\vspace{-10pt}
\end{table*}

We find that GPU usage is infrequently reported. For instance, only 16.8\% (1096 papers) and 24.7\% (1608 papers) listed their GPU type and storage information respectively. Even fewer reported information on datasets (see appendices). This is generally a serious reporting gap: only 16.51\% of papers include GPU quantity information, 24.22\% specify GPU types, and just 12.86\% report inference times. This inconsistent documentation prevents development of the structured data needed to properly understand and forecast the FM computing landscape. The absence of mandatory compute-related disclosures in author and reviewer checklists likely contributes to this gap (Figure \ref{fig:study_design} panel (B)). Conferences that enforce such requirements tend to exhibit higher reporting rates for GPU specifications.

\subsection{Self-Reported Author Survey Data}
We conduct an IRB-approved survey recruiting participants through university email lists, and social media platforms to gather a secondary source of costs associated with FM research (IRB ID: 2507001737). %: computational costs, data-related costs, and human effort. 
To incentivize participation, each respondent received a \$20 Amazon gift card, and a single first or co-first author could complete up to five surveys. The eligibility to participate was contingent on: the article focused exclusively on FM research; the paper was completed and pre-printed (publicly available online); the paper was not a survey; and the primary contribution was in the computer science domain.

% \begin{figure}[h!] % Do NOT use \begin{figure*}
% 	\centering
% 	\includegraphics[width=\textwidth]{IMG/Density_Merge.pdf} % for an image file named example_figure.*
% 	% Pick an appropriate width - in print, figures are usually one or two columns wide, which can
% 	% be approximated by 0.3\textwidth or 0.6\textwidth respectively. Use appropriate label sizes.
%         \captionsetup{font=footnotesize}
% 	% Captions go below figures
% 	\caption{\textbf{GPU Usage and TFLOPS 16 by Paper Acceptance Status and Comparison Between GPT-4o Scraped and Self-Reported Survey Data.} 
%     We compared the academia and industry affiliations' (A) GPU Number VS Average Paper Count and Average TFLOPs, (B) Average Citation VS Average GPU Number and  Average TFLOPs. 
%     }
% 	\label{$fig:density$} % give each figure a logical label name
% \end{figure}

A total of 118 institutions (267 academia first authors and 36 industry first authors) contributed to the survey. Responses were collected from 229 researchers, all of whom were first or co-first authors of publicly available papers. %In most cases, the first author was directly responsible for running the foundation model experiments and had detailed knowledge of the computing infrastructure, datasets used, and human labor involved. 
%The survey focuses on five key areas: paper information, human resource cost, computing cost, dataset cost, and other associated costs. Some of the surveyed papers may have been published in venues beyond the eight conferences included in the scraped dataset. 
Further details regarding the data collection process, both public and survey-based, as well as the analysis methods, are provided in the supplementary materials (SM).

Among the 140 papers for which both the PDF and self-reported GPU data were available, 65 first authors (46.4\%) self-reported using more GPUs than indicated in their published papers, while 21 first-authors (15.0\%) self-reported fewer GPUs. In 54 cases (38.6\%), the GPU counts reported in the papers matched the authors' self-reports. Feedback from survey respondents suggests additional GPU usage could indicate experiments that were not included in the final paper. Respondents also indicated that they often ran multiple experiments in different GPU cluster configurations, ultimately reporting only the set-up that supported the largest set of experiments. The discrepancies between the accepted scraped and survey papers are illustrated in Figure \ref{fig:study_design} panel C.

\vspace{-7pt}
\section{Results}
\vspace{-8pt}

\subsection{Foundation Model Research Is Growing in Count and Scope}
\begin{figure}[t] 
\centering\includegraphics[width=0.8\textwidth]{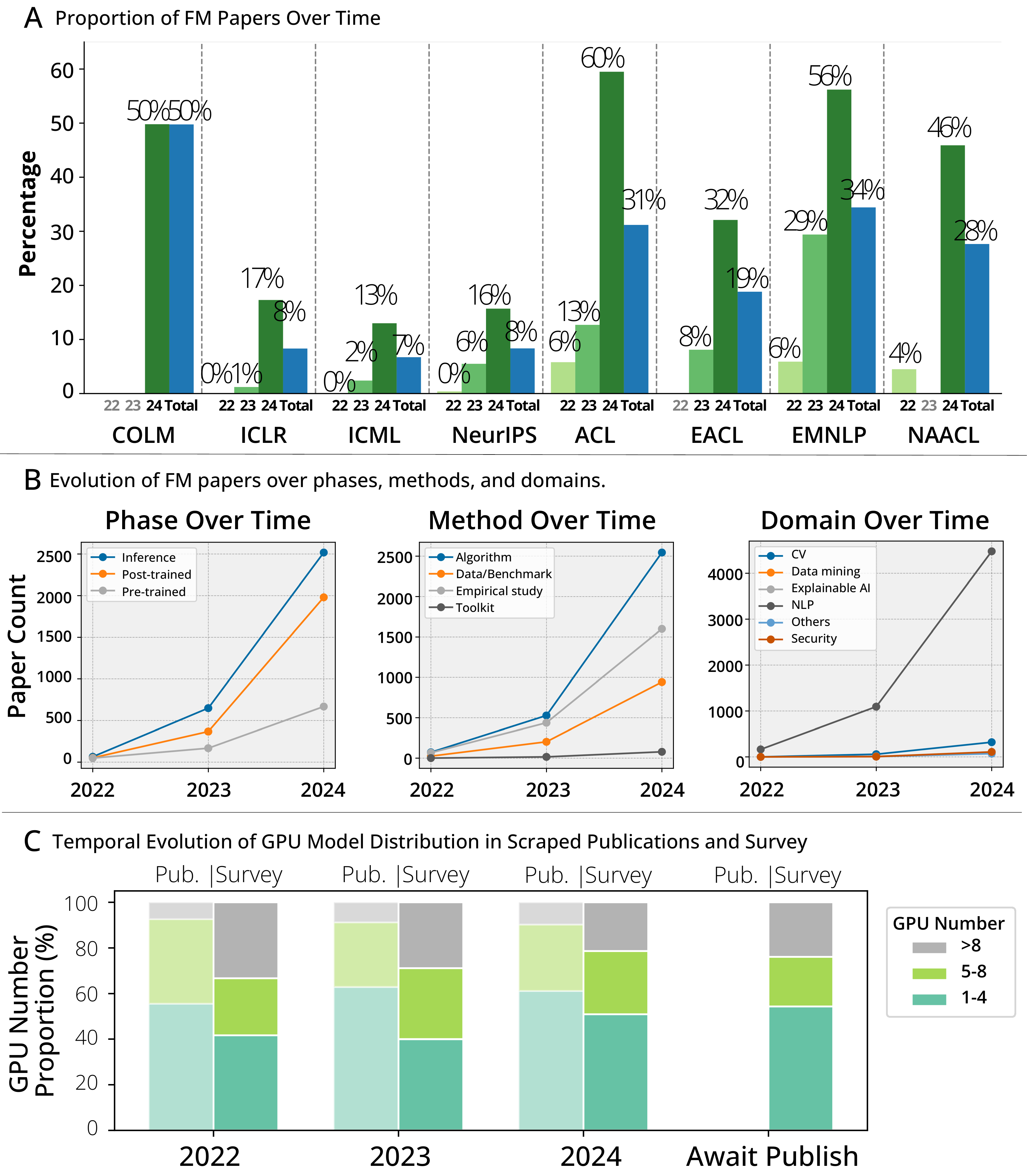}     \vspace{0.3cm}
\caption{\textbf{Temporal Evolution of FMs}. A) FM papers as a proportion of published papers over time.  
    B) Evolution of FM papers over phases, methods, and domains. 
    C) Temporal evolution of GPU Model distribution in LLM-extract and self-reported data. ``\textit{Pub.}'' denotes publications from the scraped dataset, while ``\textit{Await Publish}'' refers to papers that are either under review, rejected, or in preparation for submission.}
	\label{fig:gpu_evolution}
    \vspace{-15pt}
\end{figure}

Foundation model (FM) research has grown across phases, methods, and domains from 2022 to 2024 (Figure \ref{fig:gpu_evolution}). 
We find that FM papers grew from just 2.07\% of all accepted papers across eight key conferences in 2022 to 10.29\% in 2023, then further increased to 34.64\% in 2024 (Figure \ref{fig:gpu_evolution}A). 
FM research in natural language processing (NLP) venues specifically has grown, and conferences such as COLM, EMNLP, and ACL have the highest FM paper ratio, i.e., as compared to general ML conferences such as ICLR, ICML, NeurIPS (\cite{fan_bibliometric_2024}). 
Within FM work, inference-related papers experienced the most notable increase. Algorithmic and empirical studies outpaced other methods categories such as datasets/benchmarks and toolkits (Figure \ref{fig:gpu_evolution} B). 

Despite this upward trend, we find that GPU usage data from scraped publications and self-reported survey responses report 
steady usage (including studies currently awaiting publication) (Figure \ref{fig:gpu_evolution} C). 
Most research projects consistently use a moderate number of GPUs, typically between 1 and 8 units. Among these, the most common category remains 1 to 4 GPUs, accounting for roughly half or more of the surveyed research each year. \footnote{The reported GPU numbers commonly align with powers of two (\emph{e.g.}, 1, 2, 4, 8, and 16), likely reflecting purchasing norms for GPU clusters, which rarely involve configurations of non-standard quantities like 3, 5, 7, 11, 13, 14, or 15.}
We note that more work should be done to monitor this trend in the future, especially as the latency for GPU purchases increases.

\vspace{-8pt}
\subsection{Contributions from Both Industry and Academic Institutions}
\vspace{-8pt}
Computer science research has historically had a strong interplay between academic institutions and industrial labs (\cite{prager_research_1980}). 
We find that this still exists; academic senior authors contributed more FM papers overall, but two companies (Google and Microsoft) have the highest single-entity paper count, followed by Tsinghua University, Meta and Stanford University (Figure ~\ref{fig:country}, panels (A$_1$) and (A2$_2$)). 
Specifically, 611 academia institutions account for 4,851 papers, whereas 163 industry institution contribute 1,425 papers, alongside 239 papers affiliated with other entities such as hospitals or independent researchers. 

Importantly, the average number of publications per institution is comparable: industry-affiliated authors average 8.72 papers (SD: 29.05), while those from academia average 7.93 papers (SD: 16.62). This suggests that FM output is largely concentrated in institutions across industry and academia that can support the resource-intensive nature (\cite{mu_2024_2024}). This may also compound if resources to support large-scale model training become constrained (\cite{ahmed_growing_2023, cottier_who_2023, klyman_expanding_nodate}). 
We note that the United States and China generally lead in FM research output (Figure \ref{fig:country} panel (B)), which could be due in part to investment in higher education and AI (\cite{guest_bridging_2024}).

\subsection{FM Research is Led by Open-Weight Models}

\begin{figure}[t] 
	\centering
	\includegraphics[width=0.8\textwidth]{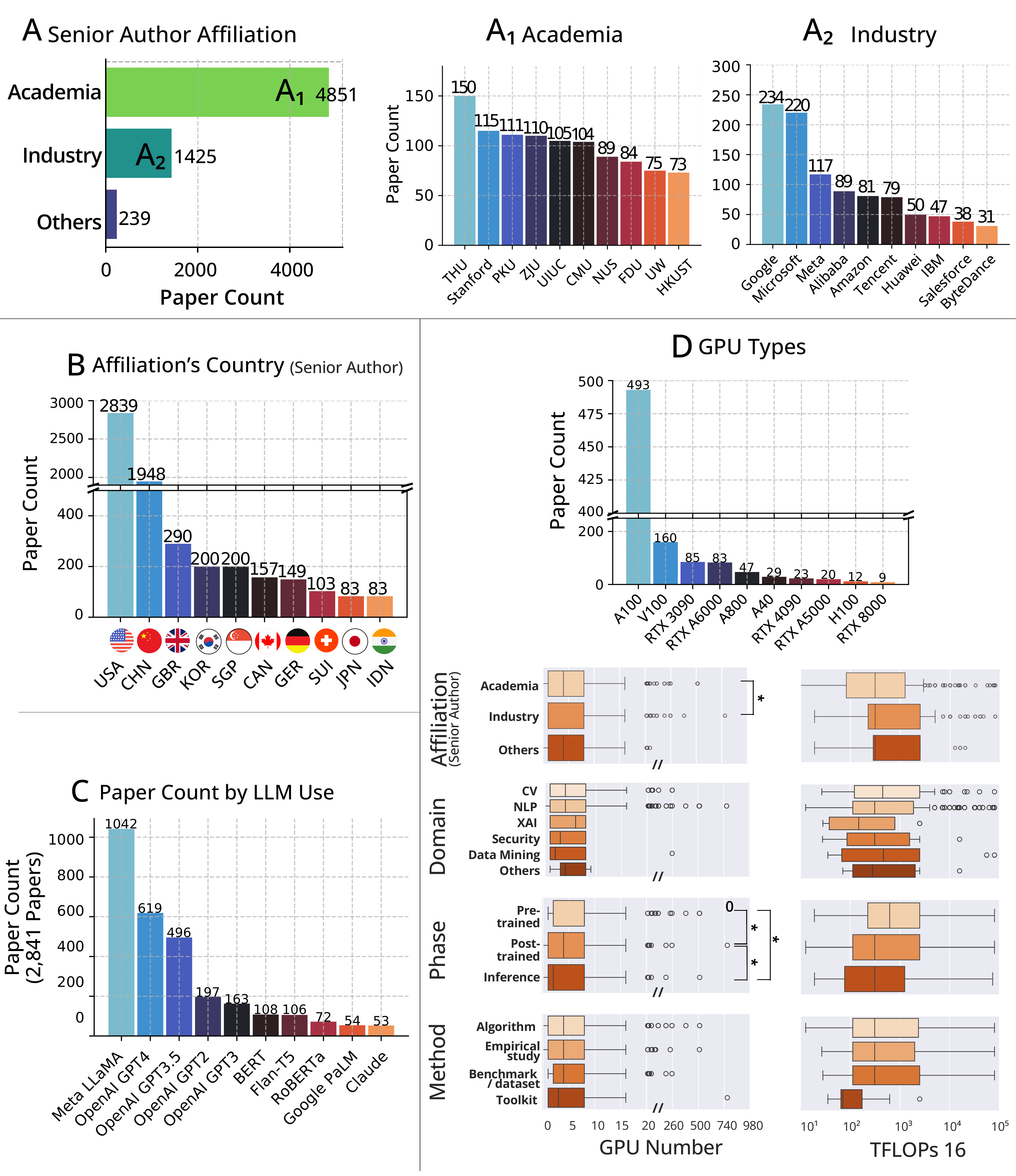}
    \vspace{0.3cm}
	\caption{\textbf{Distribution of FM Papers.} 
    We analyze FM papers across various dimensions: (A) Senior Author's Affiliations in (A1) Academia and (A2) Industry; (B) Countries by Senior Author's Affiliation; (C) Paper Count by LLM Usage; and (D) GPU Types Used. The boxplots below panel (D) display GPU Number and TFLOPs across four categories: \textit{Affiliation}, \textit{Phase}, \textit{Method}, and \textit{Domain}.
    TFLOPS: Tera Floating-Point Operations Per Second; FP16: 16-bit floating-point format (Note: * \textit{p \textless{} 0.001}).} 
	\label{fig:country} 
    \vspace{-15pt}
\end{figure}

We analyze the most frequently used model types, and find that the open-weight LLaMA models are most frequently used (Figure \ref{fig:country} panel (C)). This distinction is particularly critical in FM research. Proprietary models such as GPT maintain a significant presence, likely due to their performance and integration in commercial APIs (\cite{kapoor_position_2024}). However, open-weight models are broadly accessible, and such open-weights enable fine-tuning, domain adaptation, and benchmarking in ways that proprietary models often do not. 

\subsection{GPUs Usage in FM Research}

We investigate the distribution of specific GPU types in FM research, and find that NVIDIA\textsuperscript{\textregistered} Tesla\textsuperscript{\texttrademark} A100 cores were used most commonly (Figure \ref{fig:country} D). The top 10 GPU types are from the Nvidia family, which supports large-scale FM research. 
We compare the number of GPU and TFLOPs used in FM papers across senior author affiliation, specific application domain, phase of model development, and methodological focus. We find that both GPUs and TFLOPs are evenly distributed across most categories, with only studies focusing on pre-training reporting significantly more GPUs compared to those emphasizing post-training or inference tasks (Mann-Whitney U test, $p < 0.001$). Other variations, such as lower median GPU and TFLOP usage in FM research related to security (Figure \ref{fig:country} D - Domain) or higher usage for those constructing toolkits (Figure \ref{fig:country} D - Method), were not statistically significant.

In the set of papers analyzed, 47.4\% prioritize algorithmic development, while 31.7\% center on empirical evaluation. Only 1.5\% of the publications address toolkits or implementation infrastructure, indicating limited scholarly attention to the development of supporting tools and systems. A majority of the papers (86.4\%) focus on NLP, with a smaller portion (5.7\%) concerning CV. Regarding the stages of model development, 48.7\% of the papers investigate inference processes, 36.1\% address post-trained analysis or adaptation, and 13.3\% focus on the pre-trained stage.

%price and capabilities \cite{epoch_ai_data_2024}, high-end GPUs are more crucial in the pre-training of large models. 

\subsection{Funding Sources for FM Research}
\begin{figure}[t] 
    \centering\includegraphics[width=0.86\textwidth]{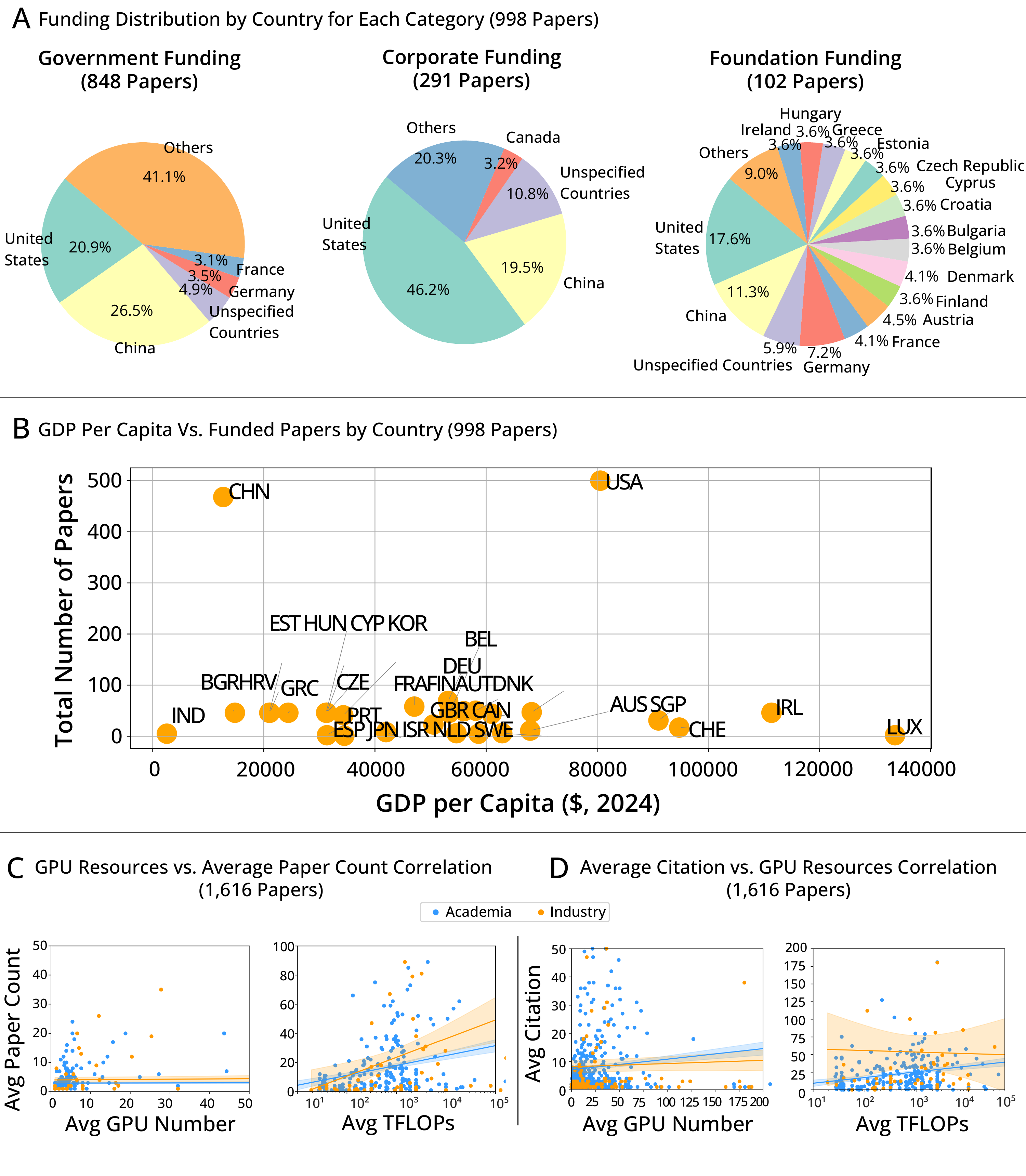}    
	\caption{\textbf{Funding Distribution of FM Research}. Only 15.3\% of FM papers contain funding country and agency information in their manuscript. A) Distribution of funding by country across three categories: Government, Corporate, and Foundation. B) Relationship between each country's GDP per capita and the number of funded papers. For academic and industry settings: C) Relationship between available GPU resources and average number of papers produced. D) Relationship between GPU resources and average citation count per paper.}
	\label{fig:gpu_funding}
    \vspace{-20pt}
\end{figure}

We investigate the sources of funding for FM research across funding types, specific agencies, and countries (Figure \ref{fig:gpu_funding}).
A total of 1126 papers had funding country information, and 998 papers had funding type information. Note that a single paper may be funded by multiple countries or organizations. 

A majority of FM research received government funding (848 papers, 85.5\%), followed by corporations (291 papers, 29.3\%), and foundations (102 papers, 10.3\%) (Figure \ref{fig:gpu_funding} A).  
We find that GDP per capita is not consistently correlated with the number of funded FM papers across countries (Figure \ref{fig:gpu_funding} B). While the U.S. and China lead in output despite differing GDP levels, FM research is closely tied to institutional support and policy than to national wealth.

\subsection{Average Effort Required}
We examine the corelation between GPU and TFLOP resources used, and both the number of papers and citations per paper (illustrated in Figure \ref{fig:gpu_funding} C \& D). We found that accepted papers typically used a median of four GPUs (mean: 46.42, SD: 932.74) and had a median of five authors (mean: 5.91, SD: 3.23). Our survey data aligned with these publications patterns, indicating a median of four actual contributors per project (mean = 4.6, SD = 3.3) and an average project duration of approximately five months (mean = 160.8 days; median = 123 days).

The GPU trend is inconsistent across both academia and industry. In contrast, when compute is measured in TFLOPs, a clearer positive relationship emerges, particularly for industry institutions (Figure \ref{fig:gpu_funding} panel (C)). Access to high-throughput compute infrastructure, rather than GPU quantity alone, plays a more decisive role in enabling sustained FM research output. The industry's dominance in the high-TFLOP range reveals the growing resource gap between private and academic institutions.

We also examine how compute resources relate to research impact, measured by citations (Figure \ref{fig:gpu_funding} panel (D)).  GPU count shows only a modest correlation with citation rates, while TFLOPs display a stronger association as institutions with higher compute tend to produce more cited work. While compute access does not guarantee impact, it does appear to confer an advantage, especially where large-scale experimentation and model deployment are feasible. But the presence of high-citation papers from less compute-intensive institutions also reveal that research influence remains multifactorial and not entirely dependent on resource scale.

Note, the average citation count for FM papers is 40.13 (median: 8.00, std: 216.02), which is comparable to other reported averages (\cite{movva_topics_2024, mohammad_examining_2020}) in high-impact fields (See Appendix Figure S8). Within FM papers, ML conferences generally have higher citations than NLP conferences, with ICML has the highest average citation (78.0, SD: 312.0).

\subsection{Accepted and Rejected Papers Resource Usage}

To date there have been few investigations into the impact of higher GPU numbers (or TFLOPs) and paper acceptance. %While many accepted papers at top venues use significant resources, particularly in deep learning and FM research, many succeed with modest setups by contributing novel ideas or theory, and some resource-intensive submissions still get rejected. Program chairs consistently report that quality, novelty and clarity of the paper, not just impressive experiments, drive acceptance decisions. 
An OpenReview analysis of 2017-2022 ICLR submissions showed review scores primarily correlated with factors such as perceived novelty and clarity (\cite{wang_what_2023}). Top conferences venues like ACL and EMNLP also instruct reviewers not to penalize papers using simpler methods (Figure \ref{fig:study_design} panel (B)), or focusing on data resources.

Our analysis of accepted and rejected ICLR papers from 2022-2024 suggests that an average rejected papers had lower GPU usage, fewer TFLOPs, and smaller author teams than accepted submissions. However, the distributions of accepted and rejected papers are very similar. 
We note that ICLR is the only conference among those studied that publicly releases rejected and withdrawn papers. This small sample size limits broader generalization to other conferences and fields.

\section{Discussion}

\paragraph{More/Better?} Our study provides empirical evidence that increasing the number of GPUs does not inherently lead to higher research impact. This is important as unchecked expansion of computational requirements further exacerbates environmental concerns (\cite{schwartz_green_2020}).
Furthermore, the current landscape of FM research remains highly centralized, with China and the United States disproportionately dominating the field, as access to computing resources often serves as a fundamental prerequisite for participation (\cite{lehdonvirta_compute_2024}).

We note that many papers utilized multiple GPUs for different tasks, making it challenging to clearly categorize GPU numbers, types, and memory configurations. Consequently, our calculated FLOPs values may differ slightly from the actual computational resources reported in these studies.

\paragraph{Open Reporting} While initiatives such as the required computing statement from some conferences acknowledge the role of computational resources, they remain insufficiently reported. Greater transparency in GPU usage (\cite{bommasani_foundation_2025}) and recognition of computing resources, including GPU availability, storage, and human labor, are integral components in evaluating AI research. This will ensure the long-term sustainability of AI research (\cite{maslej_artificial_2025}). 

While we quantified GPU usage and authorship, other resource costs are often overlooked. The cost of failed experiments is rarely acknowledged; research highlights successful outcomes, yet unsuccessful attempts are crucial to the progress of FM research. Furthermore, infrastructure costs, which vary between countries due to Gross domestic product (GDP) and AI policies, are generally not considered.

\paragraph{Automated Evaluation}
We relied on GPT-4o to extract and summarize detailed information from PDF files, a method that is susceptible to inaccuracies. We performed ten rounds of GPT-4o extraction and summarized the results through majority voting to minimize errors. Nevertheless, the extracted information may still contain inaccuracies, necessitating careful interpretation and validation.

\section*{Conclusion}

Computing resources, and GPUs in particular, have become the fuel propelling modern FM research - but they are also a source of new divides in the research community. We found that projects with access to greater GPU power generally produce more advanced pre-trained models, often achieving higher performance thanks to longer training on larger models and datasets. This advantage translates into a higher likelihood of publishing in top venues, contributing to an observed concentration of influential AI research in the hands of those with abundant compute. 

As we move into 2025 and beyond, hardware progress and growing investment will make even more GPUs available, yet the disparities in who can use vast compute may persist or even widen if deliberate steps aren't taken (\cite{muro_piloting_2024}). The future of FM research will likely feature ever-larger models and even greater compute hunger, but also a push toward more efficient methods and shared resources to counterbalance the compute divide (\cite{zhao_survey_2025, garisto_how_2024, vipra_computational_2023}). Ethically, the community is grappling with how to allocate resources in a way that is fair and sustainable, aiming to democratize AI research access while being mindful of environmental footprints. Addressing these challenges will be crucial to ensure that the next wave of AI breakthroughs is inclusive, responsible, and beneficial to all.

\clearpage

\bibliography{references}
\bibliographystyle{icml2024}

\clearpage
\appendix

\section{Methodology}
\label{app:method}
FM research is typically published in peer-reviewed computer science conferences or journals. For submissions involving experimental work, some conferences require authors to complete a checklist detailing aspects such as computational resources used and research with human subjects. After completing the initial draft, many researchers post their work on arXiv to ensure timely visibility and engagement from the community. Each submission is assigned to an area chair (AC), who invites two to three expert reviewers to evaluate the work and provide scores and feedback. Based on these reviews, the AC decides whether to accept or reject the paper, unless the authors choose to withdraw. During this review period, authors are often asked to respond to the feedback of reviewers through a rebuttal process or a revise-and-resubmit (R\&R) cycle aimed at improving the clarity and quality of the work. Acceptance rates for these conferences generally range from 25\% to 35\%.

\subsection{Public Data Collection}
In this study, we initially gathered papers related to foundation model from eight prominent peer-reviewed conferences via the OpenReview API and the Association for Computational Linguistics (ACL) Rolling Review (ARR) platform. These conferences include NeurIPS, ICLR, ICML, COLM, EMNLP, ACL, EACL, and NAACL. Every conference is held annually and only has one round; the detailed submission and publication dates are shown in Table \ref{deadlines}. The specific number of papers presented at each conference per year is shown in Table \ref{Paper_Counts}.

\begin{table}[h!]
\centering
\small
\caption{\textbf{Submission Deadlines and Conference Dates (2022-2024).} The submission deadline marks the date when researchers submit the initial version of their papers. The conference date typically corresponds to the time when accepted papers are formally published. A "-" indicates that the conference did not take place in that particular year.}
\renewcommand{\arraystretch}{1.25}
\setlength{\tabcolsep}{5pt}
\scalebox{0.88}{
\begin{tabular}{lcccccc}
\toprule
\textbf{Conference} & 
\textbf{2022 Deadline} & 
\textbf{2022 Conference Date} & 
\textbf{2023 Deadline} & 
\textbf{2023 Conference Date} & 
\textbf{2024 Deadline} & 
\textbf{2024 Conference Date} \\
\midrule
ACL & Nov.\ 15, 2021 & May 22-27, 2022 & Nov.\ 15, 2022 & July 9-14, 2023 & Nov.\ 15, 2023 & Aug.\ 11-16, 2024 \\
ICML & Jan.\ 20, 2022 & Jul.\ 17-23, 2022 & Jan.\ 26, 2023 & Jul.\ 23-29, 2023 & Jan.\ 25, 2024 & Jul.\ 21-27, 2024 \\
COLM & -- & -- & -- & -- & Mar.\ 30, 2024 & Oct.\ 7-9, 2024 \\
NeurIPS & May 19, 2022 & Nov.\ 28-Dec.\ 9, 2022 & May 17, 2023 & Dec.\ 10-16, 2023 & May 15, 2024 & Dec.\ 10-15, 2024 \\
EMNLP & Jun.\ 24, 2022 & Dec.\ 7-11, 2022 & Jun.\ 15, 2023 & Dec.\ 6-10, 2023 & Jun.\ 13, 2024 & Nov.\ 12-16, 2024 \\
EACL & Oct.\ 13, 2022 & Spring 2023 & Sep.\ 15, 2023 & -- & -- & Mar.\ 17-22, 2024 \\
ICLR & Sep.\ 28, 2022 & Apr.\ 25-29, 2022 & Sep.\ 21, 2023 & May 1-5, 2023 & Sep.\ 27, 2024 & May 7-11, 2024 \\
NAACL & Jan.\ 15, 2022 & Jul.\ 10-15, 2022 & -- & -- & Dec.\ 15, 2023 & Jun.\ 16-21, 2024 \\
\bottomrule
\end{tabular}}
\label{deadlines}
\end{table}

\begin{table}[h!]
\centering
\rowcolors{2}{white}{yellow!10}
\caption{\textbf{The proportion of selected papers among total accepted papers across eight conferences.}}
\renewcommand{\arraystretch}{1.3}
\setlength{\tabcolsep}{3pt}
\begin{tabular}{ccccccccccc}
\toprule
\textbf{Year} & \textbf{NeurIPS} & \textbf{ICLR$_\text{(Accepted)}$} & \textbf{ICML} & \textbf{COLM} & \textbf{EMNLP} & \textbf{ACL} & \textbf{NAACL} & \textbf{EACL} & \textbf{Total} & \textbf{ICLR$_\text{(Rejected)}$} \\
\midrule
2022 & 0.40\% & 0.00\% & 0.00\% & N/A & 5.90\% & 5.80\% & 4.50\% & N/A & 2.07\% & 0.00\% \\
2023 & 5.50\% & 1.21\% & 2.40\% & N/A & 29.40\% & 12.70\% & N/A & 8.10\% & 10.29\% & 0.41\% \\
2024 & 15.70\% & 17.30\% & 13.00\% & 49.80\% & 56.20\% & 59.50\% & 45.90\% & 32.10\% & 34.64\% & 11.88\% \\
\rowcolor{gray!15}
\textbf{Total} & \textbf{8.35\%} & \textbf{8.32\%} & \textbf{6.72\%} & \textbf{49.76\%} & \textbf{34.42\%} & \textbf{31.18\%} & \textbf{27.65\%} & \textbf{18.83\%} & \textbf{19.07\%} & \textbf{5.70\%} \\
\bottomrule
\end{tabular}
\label{Paper_Counts}
\end{table}

5,889 papers (16.91\%) were collected from a total of 34,828 papers. We also included 751 rejected or withdrawn papers from ICLR. After removing duplicates through paper titles (123 rejected/withdrawn papers that were resubmitted to other conferences within our dataset), we obtained 6,517 unique papers. We exclude workshop papers. The detailed list of keywords includes ``Foundation Model,'' ``Large Language Model,'' ``language model'', ``Large Model'', ``LLM,'' "GPT," "Large Vision Model," "Multi-modal Large Language Model," ``Large Vision Language Model,'' or ``VLM.''

We then employed GPT-4o mini to extract key information from the title and abstracts, including: (1) the primary domain (e.g., NLP, computer vision, data mining, security, explainable AI, or other specialized domains like systems or HCI); (2) the primary phase of the paper (e.g., pre-trained, post-trained, or inference); and (3) the primary methodology (e.g., algorithm, dataset/benchmark, empirical study, or toolkit). We independently ran GPT-4o mini to analyze each paper's title and abstract five times and determined its final domain, phase, and methodology through majority voting. In cases where multiple domains, phases, or methods were present, we selected only the first mentioned, treating it as the primary category.

We used OpenAI's GPT-4o mini to process all the PDF documents through an integrated OCR (Optical Character Recognition) pipeline that converts text content into machine-readable text before applying language model analysis. The system skips the parts that contain images. Once the text is extracted, GPT-4o mini parses the document in segments, using contextual understanding to identify relevant sections. The process is optimized for robustness and generalization across heterogeneous formats, enabling high-quality downstream information extraction, such as identifying GPU specifications, datasets, and/or funding disclosures from papers.
We employed an automated service SerpAPI to collect paper citation data from Google Scholar using the paper's title, up to the cutoff date of March 11th, 2025.

Here are the details of the conference's full name, acceptance rate, and h5 index *:
\begin{enumerate}
    \item NeurIPS: Conference on Neural Information Processing Systems (acceptance: h5 index = 337)
    \item ICLR: International Conference on Learning Representations (h5 index = 304)
    \item ICML: International Conference on Machine Learning (h5 index = 268)
    \item COLM: Conference on Language Modeling (h5 index = N/A)
    \item EMNLP: Empirical Methods in Natural Language Processing (h5 index = 193)
    \item ACL: Meeting of the Association for Computational Linguistics (h5 index = 215)
    \item EACL: Conference of the European Chapter of the Association for Computational Linguistics (h5 index = 56)
    \item NAACL: Conference of the North American Chapter of the Association for Computational Linguistics (h5 index = 132)

\end{enumerate}
* h5 index: the h-index for articles published in the last 5 complete years from Google Scholar

\begin{figure}[h!] 
	\centering
	\includegraphics[width=0.6\textwidth]{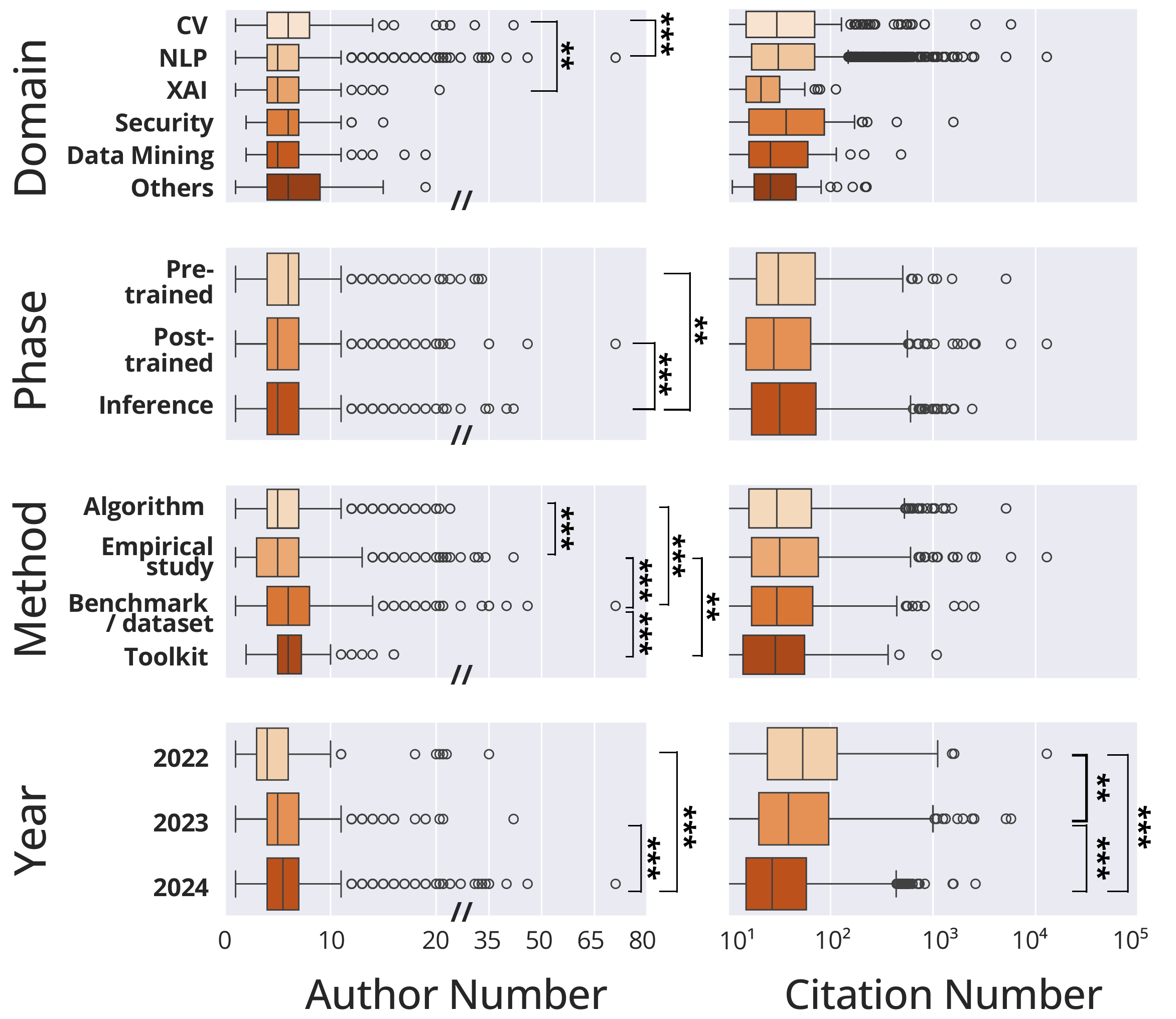}
    \vspace{0.3cm}
	\caption{\textbf{Author Number and Citation Number Versus Senior Authors' Affiliation, Domain, Phase, and Methodology.} (Note: * \textit{p \textless{} 0.001}).} 
	\label{fig:append_Boxplots} 
    \vspace{-15pt}
\end{figure}

% \begin{figure}[t] 
% 	\centering
% 	\includegraphics[width=0.6\textwidth]{IMG/MIT_CALC Appendix Density.pdf}
%     \vspace{0.3cm}
% 	\caption{\textbf{Accepted Versus Rejected ICLR Conferences GPU Distribution.}}
% 	\label{fig:append_density} 
%     \vspace{-15pt}
% \end{figure}

% \begin{figure}[t] 
% 	\centering
% 	\includegraphics[width=0.3\textwidth]{IMG/MIT_CALC Appendix CitationVsYear.pdf}
%     \vspace{0.3cm}
% 	\caption{\textbf{Author Number and Citation Number Versus Senior Authors' Affiliation, Domain, Phase, and Methodology.} (Note: * \textit{p \textless{} 0.001}).} 
% 	\label{fig:append_citation+years} 
%     \vspace{-15pt}
% \end{figure}

\subsubsection{GPT-4o mini Prompt Template}

\begin{promptbox}{GPT-4o Prompt for abstract information scraping}
We would like to summarize this abstract. If you don't know the answer, output "nan".
Please answer these questions. Only return JSON as a dictionary:
"key (only number)" + "value (list) ([in-context reasoning, final answer])", no commentary:

[Output a list] Select one primary domain from the following options:
NLP, CV, Data Mining, Security, Explainable AI, or others (please specify such as system, HCI, ).

[Output a list] Select one phase that this project experiments primarily include:
pre-trained, post-trained, and inference.

[Output a list] Primary Method: algorithm, dataset/benchmark, empirical, toolkit.
Here is the definition: An algorithm is a step-by-step process for solving a problem
or completing a task through computational instructions. A dataset is a collection
of data used for training or testing models, while a benchmark is a standard dataset
used to evaluate performance. Empirical refers to knowledge gained from real-world
observations or experiments, providing evidence for or against a method. A toolkit
is a set of software tools or libraries designed to aid developers in tasks like
model training and deployment, such as TensorFlow or Scikit-learn.
\end{promptbox}

\begin{promptbox}{GPT-4o Prompt for Extracting Information from PDFs}
We would like to summarize from this pdf. Calculate all the cost in USD.
If you don't know the answer, output "nan".
Please answer these questions with REASONING which contains evidence from the given text.
Only return JSON as a dictionary: "key(only number)" + "value (list) (in-context reasoning + final answer)", no commentary:

1. [Output a list] Select one primary domain from the following options:
   NLP, CV, Data Mining, Security, Explainable AI, or others (please specify such as system, HCI, ).
   
2. [Output a list] Select one phase that this project experiments primarily include:
   pre-trained, post-trained, and inference.
   
3. [Output a list] Primary Method: algorithm, dataset/benchmark, empirical, toolkit.

4. [Output a string] Number of GPU(s) used.

5. [Output a string] Type of GPU used:
   NVIDIA Tesla V100, NVIDIA Tesla A100, NVIDIA Tesla P100, NVIDIA Tesla A40,
   NVIDIA Tesla A800, NVIDIA Tesla A4000, NVIDIA Tesla A5000, 3090, 4090.
   
6. [Output a string] GPU Storage: 16GB, 24GB, 32GB, 40GB, 48GB, 80GB, 96GB.

7. [Output a string] Inference time: Retrieve the original key context or description
   that contains GPU time (<50 words).
   
8. [Output a list] LLM(s) used for fine-tuning or inference in the paper:
   OpenAI GPT4o, OpenAI GPT4, OpenAI GPT3.5, OpenAI GPT3, OpenAI GPT2,
   Google Gemini, Google Bert, Meta LLaMA, Google PaLM, Anthropic Claude,
   Megatron-Turing, Flan-T5, others (please specify).
   
9. [Output a list] LLM(s) used for evaluation in the paper:
   OpenAI GPT4o, OpenAI GPT4, OpenAI GPT3.5, OpenAI GPT3, OpenAI GPT2,
   Google Gemini, Google Bert, Meta LLaMA, Google PaLM, Anthropic Claude,
   Megatron-Turing, Flan-T5, others (please specify).
   
10. [Output a list] GitHub Link ONLY for this paper.

11. [Output a list] Funding Resource Region/Country (ONLY search in acknowledgment):
    China, US, EU, Canada, Australia, Others.
    
12. [Output a list] Funding Resource (ONLY search in acknowledgment):
    Salary, Government, Corporate, NGO, Foundation.
    
13. [Output a string] Dataset Curation/Construction Cost.

14. [Output a string] API Cost (i.e. model testing, and iterative retraining).

15. [Output a string] Estimate \# of Rows.

16. [Output a string] Average Length of each data item (tokens).

17. [Output a string] Data Storage.

18. [Output a list] Other type(s) and amount(s) of cost mentioned in the paper.

    If yes, output the structure like this:
    
    1. Cost source 1 (selected one from the sample list: hardware devices, cloud services,
       software licensing fee, participant incentives, energy/environmental costs,
       retraining costs, others)
       
    2. Total \$ of Cost source 1
    
    3. Cost source 2 (...)
    
    4. Total \$ of Cost source 2
    
    5. Cost source 3 (...)
    
    6. Total \$ of Cost source 3
\end{promptbox}

\section{Survey Questions}

The survey study's IRB is approved by MIT Committee On the Use of Humans as Experimental Subjects (COUHES). Data collection took place from September 2024 to December 2024.
Several major machine learning conferences require authors to report, and reviewers to consider, information related to computing resources. These checklist items are intended to promote responsible machine learning research by encouraging documentation of reproducibility, transparency, ethical considerations, and potential societal impact. 

We examined the author and reviewer guidelines of eight conferences and incorporated relevant items into our survey (\autoref{tab:conference_checklist}). Notably, the four ARR conferences (EMNLP, ACL, NAACL, and EACL) use a unified checklist and review form, which include more detailed questions on computing resources than the other conferences. Our survey was designed to reflect this specificity, including questions on GPU usage, datasets, and human labor involved in the research process.  

Survey responses were reviewed for eligibility and quality by Y. Hao, Y. Huang, and C. Zhao. A total of 56 responses were excluded due to ineligibility, including submissions from non-first authors, duplicate entries from co-first authors on the same paper, or unpublished or incomplete work. Additional exclusions were made for responses unrelated to LLM research or those deemed invalid due to unanswered mandatory questions or illogical answers.

Following data cleaning, we cross-referenced the paper titles and their corresponding ArXiv/OpenReview/PDF links with the papers in our public data collection database. A total of 122 papers were already included in the database. Since the survey was open to researchers beyond the eight peer-reviewed conferences initially targeted, we received papers from 15 different conferences and journals. For papers available on OpenReview with review scores and context, we followed the same process as outlined in \autoref{app:method}. For papers available only via ArXiv or PDF links, we used GPT-4o mini to extract relevant context, including computational resources, dataset details, and funding information.

The detailed survey items are as follows:

\subsection*{Part I: Paper Information}
\begin{itemize}
  \item Contact Email for Gift Card
  \item Paper Name (Please insert the completed published title which can be retrieved through Google Scholar) \textasteriskcentered
  \item Paper OpenReview/ArXiv PDF Link
  \item Published Year \textasteriskcentered
  \item If published: I published at conference/journal name ...... (i.e., ICML 2024 Poster; ECCV 2023 Workshop)
\end{itemize}

\subsection*{Part II: Human Resource Cost Information}
\begin{itemize}
  \item Project Start Month Year (i.e. June, 2023) \textasteriskcentered
  \item Project Completion Month Year (i.e. May, 2024) \textasteriskcentered
  \item Preprint Month, Year (i.e. August, 2024) \textasteriskcentered
  \item Numbers of paper contributors (who actually work/provide guidance on the project, not just the manuscript) \emph{NUMBERS ONLY} \textasteriskcentered
  \item How many times you submit this paper and get accepted? \textasteriskcentered
  \item Funding Resource Region/Country(s)
  \item Funding Resource Type(s)
\end{itemize}

\subsection*{Part III: Computing Cost Information}
\begin{itemize}
  \item My paper used: GPU/CPU/didn't use any computing resource \textasteriskcentered
  \item Number of GPUs that I used in this paper: \emph{NUMBERS ONLY}
  \item Types of GPU that I am using for this paper
  \item My GPU's Storage Unit (Each)
  \item Average Single GPU running time: (estimate in HOURS)
\end{itemize}

\subsection*{Part IV: Dataset Cost Information}
\begin{itemize}
  \item Dataset Platform used for data collection, cleaning, labeling, evaluation, etc \textasteriskcentered
  \item Dataset Total Cost used for data collection, cleaning, labeling, evaluation, etc (in USD, \emph{NUMBERS ONLY}) \textasteriskcentered
  \item API Cost in USD (i.e. model testing, and iterative retraining): \emph{NUMBERS ONLY}
  \item Data Storage (in MB): \emph{NUMBERS ONLY}
\end{itemize}

\subsection*{Part V: Other Information}
\begin{itemize}
  \item Other costs not mentioned in this survey? For example, wasted cost because of wrong temperature and inference time?
  \item Others Comments
  \item Feedback on this Survey
\end{itemize}

\smallskip
\noindent\textit{* indicates required}

\begin{table}[t]
    \centering
    \scriptsize
    \begin{threeparttable}
    \caption{\textbf{Conferences and Computing Resource Related Checklist.}}
    \renewcommand{\arraystretch}{2}
    \begin{tabular}{p{1.5cm}p{1.3cm}p{7.5cm}p{4.5cm}}
    \toprule[1pt]
        \textbf{Conferences} & \textbf{Computing Resource Related Checklist?} & \textbf{Specific Items} & \textbf{Should Reviewers Consider Computing Resources in Publication Decisions?} \\
        \hline
        NeurIPS &  NeurIPS Paper Checklist Guidelines &
        Experiments Compute Resource: For each experiment, does the paper provide sufficient information on the computer resources (type of compute workers, memory, time of execution) needed to reproduce the experiments?
        1) The answer NA means that the paper does not include experiments.
        2) The paper should indicate the type of compute workers, CPU or GPU, internal cluster, or cloud provider, including relevant memory and storage.
        3) The paper should provide the amount of compute required for each of the individual experimental runs as well as estimate the total compute.
        The paper should disclose whether the full research project required more compute than the experiments reported in the paper (e.g., preliminary or failed experiments that didn't make it into the paper)
        & NeurIPS Reviewer Guidelines\tnote{a}: Not specifically mentioned\\
        \hline

        ICLR & No & Not required but can mention in the Reproducibility Statement &
        ICLR Reviewer Guide\tnote{b}: Not specifically mentioned. \\
        \hline

        ICML & ICML Paper Guidelines &
        If you ran experiments, did you include the amount of compute and the type of resources used (e.g., type of GPUs, internal cluster, or cloud provider)? &
        ICML 2025 Reviewer Instructions\tnote{c}: Not specifically mentioned. Did you review the supplementary material? Which parts?\\
        \hline

        COLM &  No &
        Not required but can be mentioned in the Reproducibility Statement &
        COLM Review Guidelines\tnote{d}: ...we ask that you take into account that most researchers do not have access to large-scale compute. While some questions in our field require significant computational resources to study at the upper limits of scale, few publishing authors have these resources. Limiting this type of research to only these labs will stifle innovation and understanding. Naturally, this runs the risk that some small-scale results will not hold when studied later on at a large scale. But some results will, and they will not make it unless we, the program committee, make a bet on them. \\
        \hline

        EMNLP, ACL, NAACL, EACL & {ARR} &
        Did you run computational experiments? 
        
        - Did you report the number of parameters in the models used, the total computational budget (e.g., GPU hours), and the computing infrastructure used? 
        
        Did you use human annotators (e.g., crowdworkers) or research with human subjects? 
        
        - Did you report the full text of instructions given to participants, including e.g., screenshots, disclaimers of any risks to participants or annotators, etc.? 
        
        - Did you report information about how you recruited (e.g., crowdsourcing platform, students) and paid participants, and discuss if such payment is adequate given the participants' demographic (e.g., country of residence)?
         &
        Review Form\tnote{e}: Is there enough information in this paper for a reader to reproduce the main results, use results presented in this paper in future work (e.g., as a baseline), or build upon this work? If the authors state (in anonymous fashion) that datasets will be released, how valuable will they be to others? \\
    \bottomrule[1pt]
    \end{tabular}

    \begin{tablenotes}
        \scriptsize
        \item[a] \url{https://neurips.cc/Conferences/2024/ReviewerGuidelines}
        \item[b] \url{https://iclr.cc/Conferences/2025/ReviewerGuide}
        \item[c] \url{https://icml.cc/Conferences/2025/ReviewerInstructions}
        \item[d] \url{https://colmweb.org/ReviewGuide.html}
        \item[e] \url{https://aclrollingreview.org/reviewform}
    \end{tablenotes}
    \label{tab:conference_checklist}
    \end{threeparttable}
\end{table}

\end{document}